\documentclass[lettersize,journal]{IEEEtran}
\usepackage{amsmath,amsfonts}
\usepackage{color}
\usepackage[caption=false,font=normalsize,labelfont=sf,textfont=sf]{subfig}
\usepackage{textcomp}
\usepackage{stfloats}
\usepackage{url}
\usepackage{times}
\usepackage{verbatim}
\usepackage{graphicx}
\usepackage[nocompress]{cite}
\usepackage{longtable}
\usepackage[utf8]{inputenc}
\usepackage[T1]{fontenc}
\usepackage{lipsum}
\usepackage[left=.6in,right=.6in,top=1.2 in,bottom=1.2 in]{geometry}

\usepackage{algorithm}
\usepackage{algorithmic}
\begin{document}
\makeatletter
\newcommand{\nosemic}{\renewcommand{\@endalgocfline}{\relax}}
\newcommand{\dosemic}{\renewcommand{\@endalgocfline}{\algocf@endline}}
\newcommand{\pushline}{\Indp}
\newcommand{\popline}{\Indm\dosemic}
\let\oldnl\nl
\newcommand{\nonl}{\renewcommand{\nl}{\let\nl\oldnl}}
\makeatother
\title{Machine Learning-Based AP Selection in User-Centric Cell-free Multiple-Antenna Networks}

\author{\IEEEauthorblockN{Shirin Salehi\IEEEauthorrefmark{1}, Saeed Mashdour\IEEEauthorrefmark{2}, Orhun Tamyigit\IEEEauthorrefmark{1}, Sadra Seyedmasoumian\IEEEauthorrefmark{1}, Majid Moradikia\IEEEauthorrefmark{3},
Rodrigo C. de Lamare\IEEEauthorrefmark{2}\IEEEauthorrefmark{4} and Anke Schmeink\IEEEauthorrefmark{1}
}

\IEEEauthorblockA{\IEEEauthorrefmark{1} Chair of Information
Theory and Data Analytics (INDA), RWTH Aachen University, Aachen, Germany}

\IEEEauthorblockA{\IEEEauthorrefmark{2} Centre for Telecommunications Studies, Pontifical Catholic University of Rio de Janeiro, Rio de Janeiro 22541-041, Brazil}    

\IEEEauthorblockA{\IEEEauthorrefmark{3} Department of Data Science, Worcester Polytechnic Institute (WPI), 100 Institute
Road, Worcester, MA, 01609-2280, USA}    

\IEEEauthorblockA{\IEEEauthorrefmark{4} School of Physics, Engineering and Technology, University of York, United Kingdom}


    \{shirin.salehi, sadra, anke.schmeink\}@inda.rwth-aachen.de, orhun.tamyigit@rwth-aachen.de, 
    smashdour@gmail.com, majidemoradikia@gmail.com, delamare@puc-rio.br \vspace{-3mm}
}


\maketitle

\begin{abstract}
User-centric cell-free (UCCF) massive multiple-input
multiple-output (MIMO) systems are considered a viable solution to realize the advantages offered by cell-free (CF) networks, including reduced interference and consistent quality of service while maintaining manageable complexity. In this paper, we propose novel learning-based access point (AP) selection schemes tailored for UCCF massive MIMO systems. The learning model exploits the dataset generated from two distinct AP selection schemes, based on large-scale fading (LSF) coefficients and the sum-rate coefficients, respectively. The proposed learning-based AP selection schemes could be implemented centralized or distributed, with the aim of performing AP selection efficiently. We evaluate our model's performance against CF and two heuristic clustering schemes for UCCF networks. The results demonstrate that the learning-based approach achieves a comparable sum-rate performance to that of competing techniques for UCCF networks, while significantly reducing computational complexity.
\end{abstract}

\vspace{-0.75em}

\section{Introduction}
Cell-free (CF) massive multiple-input multiple-output (MIMO) ~\cite{ngo2015cell, ngo2017cell,itprec,rmmsecf,rscf,cscf,csidd,iddllr,rrscf} is a potential architecture to meet future performance demands of B5G/6G that extends the initial success of massive MIMO systems and centralized large-scale MIMO concepts \cite{mmimo,wence,xlanzheng} such as consistent high data rates everywhere, uniform quality of service (QoS), ultra-high reliability, and avoidance of cell interference, as the concept of cell boundaries is nonexistent in this approach~\cite{kassam2023review}. In a CF network, a large number of access points (APs) serve a smaller number of user equipments (UEs) over the same time-frequency resources. 

However, employing the ideal CF network, where each UE is served by all APs, is impractical due to limited fronthaul capacity and finite computational resources. Clustering with a user-centric cell-free (UCCF) approach is suggested in the literature to leverage the benefits of CF networks while maintaining manageable complexity~\cite{ammar2021user, bjornson2020scalable, demir2021foundations,tentu2022uav, mashdour2022enhanced, guevara2021partial, wei2022user,cscf}, where each user is served by a personalized cluster of nearby APs. 

For example, in~\cite{bjornson2020scalable}, the authors define a scalable CF massive MIMO system and design a scalable algorithm for joint initial access, pilot assignment, and cooperation cluster formation. The scalability is defined in terms of the number of UEs, $K$, and ensures finite complexity and resource requirements for each AP as $K \rightarrow \infty$. In this work, the UE appoints the AP with the strongest large-scale fading (LSF) coefficient as its master AP. The master AP then assigns a pilot with minimum pilot contamination to the UE. Finally, each neighboring AP decides to serve this UE if it does not serve any UEs on that pilot or if the new UE has a better channel condition than the one that is currently being served. In another work~\cite{dao2020effective}, an AP selection algorithm is suggested based on two newly proposed metrics. These metrics measure the channel quality and effective channel gain between every UE and AP and are merely based on LSF coefficients.

However, the clustering problem is highly combinatorial and can be prohibitively time-consuming to solve optimally when UEs are in motion. Deep neural networks (DNNs) are promising candidates for performing real-time AP selection in a dynamic environment where UEs are in motion and AP selection needs to be frequently updated. Moreover, a distributed DNN approach could also alleviate the significant fronthaul and computational load. For instance in~\cite{ranasinghe2021graph}, a graph neural network-based AP selection is proposed for CF massive MIMO systems, which consists of two graphs: a homogeneous graph to represent the structure of APs and a heterogeneous one to represent both APs and UEs. The main achievement is that it requires a small number of RSRP measurements. However, the model relies on the leveraged knowledge of the statistic correlated value map and AP placement to predict the potential links which makes the model complex and limits its scalability in terms of the number of APs.   



In this paper, we introduce two learning models based on feedforward DNNs to approximate access point (AP) selection schemes introduced in our previous work~\cite{mashdour2024clustering}. These schemes establish thresholds on LSF and sum-rate criteria, aiming to maximize the overall network sum-rate. In contrast to the existing literature which often relies on feature vectors composed of LSF coefficients for training, we also utilize sum-rates to integrate small-scale fading into the decision-making process. Hence, the model based on sum-rate feature vector is able to differentiate across different antennas of a single AP, resulting in a sum-rate performance that closely aligns with that of a cell-free network with no clustering. Another distinguishing aspect of our approach is its user-centric decision-making, as opposed to the network-centric approach commonly found in the literature. These models can operate in either a centralized or distributed fashion. In the centralized approach, computations are performed at a CPU, while in the distributed approach, computations occur at each UE.

A key advantage of employing learning-based models is the significant reduction in computation time required for AP selection. Additionally, the distributed schemes offer notable scalability, accommodating scenarios involving UE mobility or varying numbers of UEs. In such dynamic environments, where computations would otherwise need to be recalculated from scratch, the efficiency and adaptability of learning-based methods become particularly invaluable.

This paper is structured as follows: Section II presents the system model including the CF and UCCF. The clustering schemes based on LSF coefficients and sum-rate are presented in Section III. Section IV presents the learning-based AP selection schemes. Section V presents and discusses the simulation results, whereas Section VI draws the conclusions.

\section{System Model}

We investigate the downlink scenario of a cell-free massive MIMO (CF-mMIMO) network, consisting of $L$ APs, each equipped with $N$ uniformly spaced antennas. Additionally, there are $K$ single-antenna UEs randomly distributed across the network. The total number of AP antennas, denoted as $M = LN$, is assumed to be greater than the number of UEs, i.e., $M > K$.

The channel coefficient that connects the $m$th antenna of an AP to the $k$th UE is denoted by $g_{m,k} = \sqrt{\beta_{m,k}} h_{m,k}$, where $\beta_{m,k}$ represents the large-scale fading coefficient and $h_{m,k} \sim \mathcal{CN}(0, 1)$ symbolizes the small-scale fading coefficient. Small-scale fading coefficients are independent and identically distributed (i.i.d.) random variables (RVs), consistent over a coherence interval and independent across different intervals.

The downlink signal in a CF network can then be expressed as:
\begin{equation}
\mathbf{y}_{CF} = \sqrt{\rho_f} \mathbf{G}^T \mathbf{P} \mathbf{x} + \mathbf{w},
\end{equation}
where $\rho_f$ indicates the peak transmit power per antenna. The channel matrix $\mathbf{G} = \hat{\mathbf{G}} + \tilde{\mathbf{G}} \in \mathbb{C}^{LN \times K}$ includes both the estimated channel $\hat{\mathbf{G}}$, and the estimation error $\tilde{\mathbf{G}}$ which accounts for the imperfections in channel state information (CSI). The matrix entries are indicated by $\left[\mathbf{G}\right]_{m,k} = g_{mk}$. Other components are the precoding matrix $\mathbf{P} \in \mathbb{C}^{LN \times K}$, alongside the symbol vector $\mathbf{x} = \left[x_1, \cdots, x_K\right]^T$, where $\mathbf{x} \sim \mathcal{CN}(\mathbf{0}, \mathbf{I}_K)$, and the noise vector $\mathbf{w} = \left[w_1, \cdots, w_K\right]^T$, with $\mathbf{w} \sim \mathcal{CN}(0, \sigma_w^2 \mathbf{I}_K)$. Given that the signal and noise are Gaussian, and all elements of $\mathbf{x}$ are statistically independent from the noise and channel coefficients, we can calculate the sum-rate of the CF system as follows:
\begin{equation}\label{eq:SR_CF}
SR_{CF} = \log_2\left(\det\left[\mathbf{R}_{CF} + \mathbf{I}_K\right]\right),
\end{equation}
where $\mathbf{R}_{CF}$, the covariance matrix, is defined by:
\begin{equation}\label{eq:R_CF}
\mathbf{R}_{CF} = \rho_f \hat{\mathbf{G}}^T \mathbf{P} \mathbf{P}^H \hat{\mathbf{G}}^* \left(\rho_f \tilde{\mathbf{G}}^T \mathbf{P} \mathbf{P}^H \tilde{\mathbf{G}}^* + \sigma_w^2 \mathbf{I}_K\right)^{-1}.
\end{equation}

\begin{figure}
    \centering
    \includegraphics[width=.8\linewidth]{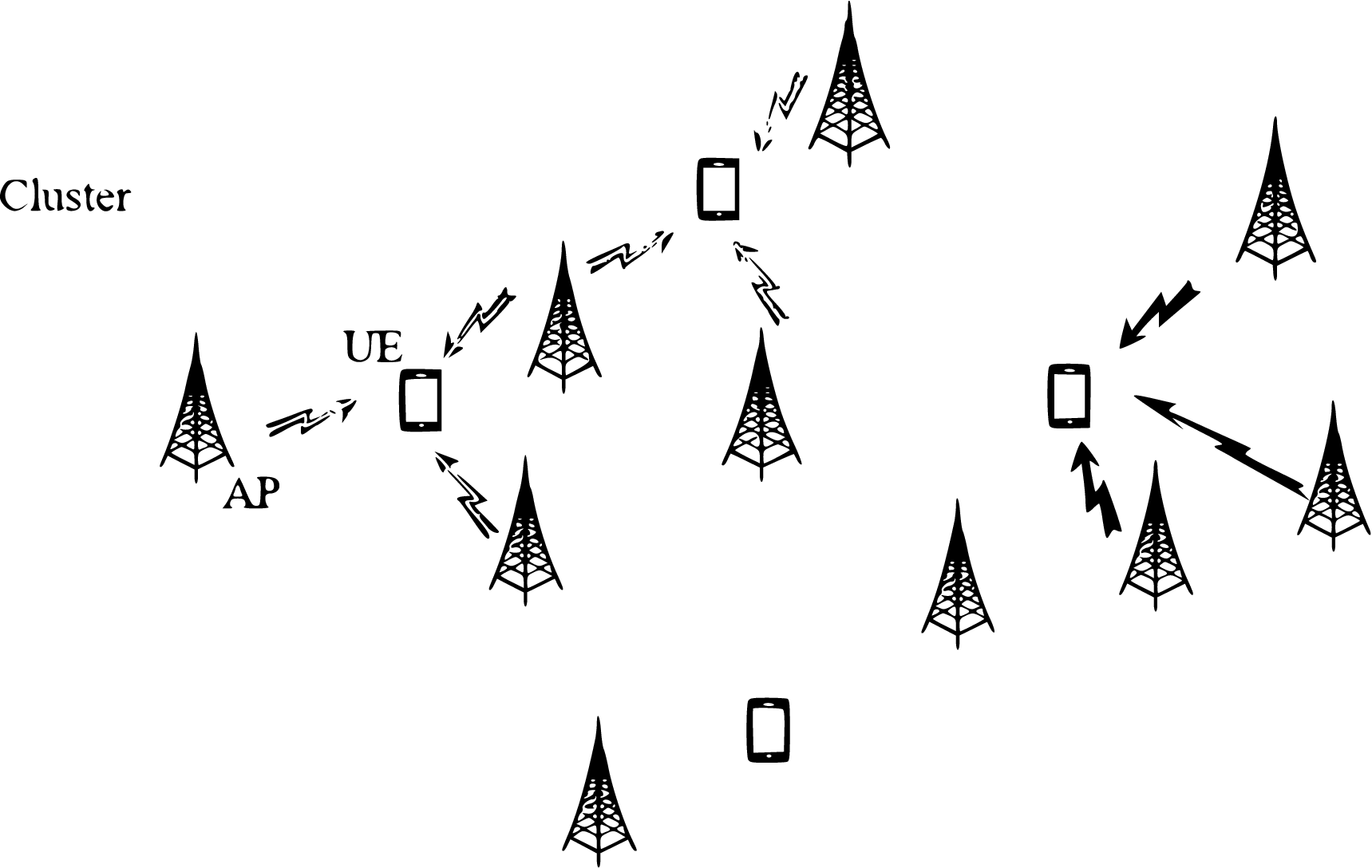}
    \vspace{-1em}
    \caption{UCCF clustered network. 
    }
    \vspace{-1em}
    \label{UCCF}
\end{figure}

For the UCCF massive MIMO network illustrated in Fig.~\ref{UCCF},
 the received downlink signal is
\begin{equation} \label{eq:CF-sig}
    \textbf{y}_{UC}=\sqrt{\rho _{f}}\textbf{G}_{a}^T\textbf{P}_a\textbf{x}+\textbf{w}.
\end{equation}
Similar to the CF network, the sum-rate of the UCCF system is given by
\begin{equation}\label{eq:RCF}
    SR_{UC}=\log_{2}\left (  \det\left [\textbf{R}_{UC}+\textbf{I}_K  \right ]\right ),
\end{equation}
where the matrix $\textbf{R}_{UC}$ is
\begin{equation}\label{eq:RCF_1}
    \textbf{R}_{UC}=\rho _{f} \hat{\textbf{G}}_a^{T}\textbf{P}_{a}\textbf{P}_{a}^{H}\hat{\textbf{G}}_a^{\ast }\left ( \rho _{f}\tilde{\textbf{G}}_a^{T}\textbf{P}_{a}\textbf{P}_{a}^{H}\tilde{\textbf{G}}_a^{\ast } +\sigma _{w}^{2}\textbf{I}_K\right )^{-1},
\end{equation} 
and the channel and precoder matrices in this network are considered as $\textbf{G}_{a}=\left [ \textbf{g}_{a1},\cdots , \textbf{g}_{aK}\right ]$ and $\textbf{P}_{a}$ as a function of $\textbf{G}_{a}$,  respectively, where $\textbf{g}_{ak} \in \mathbb{C}^{{LN}\times 1}$, $k \in \left \{ 1,\cdots ,K \right \}$. We define  $\textbf{g}_{ak}=\textbf{A}_{k}\textbf{g}_{k}$ where $\textbf{A}_{k}=\textup{diag}\left ( a_{k1},\cdots ,a_{k{LN}} \right )$ are diagonal matrices with entries given by
\begin{equation} \label{akl}
    a_{kl}=\left\{\begin{matrix}
1 & if \  l\in U_{k}\\ 
0 & if \  l\notin  U_{k}
\end{matrix}\right.
, l\in \left \{ 1,\cdots ,{LN} \right \},
\end{equation}
where $U_{k}$ denotes the subset of APs serving $k$th UE, $\textbf{g}_{k} \in \mathbb{C}^{{LN}\times 1}$, $k \in \left \{ 1,\cdots ,K \right \}$ is the $k$th column of the ${LN}\times K$ CF channel matrix $\textbf{G}=\left [ \textbf{g}_{1},\cdots , \textbf{g}_{K}\right ]$. Several alternative designs can be employed for precoding \cite{wence,grbd,mbthp,rmmsecf,rrscf} at the transmit side, i.e., at the AP level, and interference mitigation \cite{jidf,sjidf,spa,mfsic,mbdf,bfidd,csidd,iddllr} at the UEs. 

\section{Clustering Methods in UCCF network}

The selection of APs for serving UEs in UCCF networks, known as clustering, can be based on various criteria. In this section, we explore two distinct clustering methodologies for AP clustering. The first method utilizes the LSF criterion to determine optimal clustering. The second method, recently introduced in \cite{mashdour2024clustering}, employs an information rate criterion known as Boosted sum-rate (BSR) which aims to enhance AP clustering. Both methods aim to maximize the sum-rate of the UCCF network heuristically, ideally achieving a sum-rate comparable to that of the corresponding cell-free (CF) network.


\subsection{LSF Based Clustering}
The clustering approach using the LSF criterion is founded on large-scale propagation characteristics such as path loss and shadowing to select appropriate APs for UEs. The LSF method emphasizes the general attenuation of signals over distances but overlooks specific aspects of the wireless environment.

In this technique, an AP $m$ is chosen for clustering with a user $k$ if the average channel gain exceeds a predefined threshold $\alpha_{lsf}$, such that $\beta_{m,k} \geq \alpha_{lsf}$. If no AP meets this criterion for a given user, the AP that offers the highest average channel gain to that user is selected as part of the cluster set, as demonstrated in the following equation:
\vspace{-2mm}
\begin{equation}
    U_{k,lsf} = \left\{ m : \beta_{m,k} \geq \alpha_{lsf} \right\} \cup \left\{ \underset{m}{\text{argmax}} \ \beta_{m,k} \right\}
\end{equation}
The threshold $\alpha_{lsf}$ is defined as:
\vspace{-2mm}
\begin{equation}
    \alpha_{lsf} = \frac{1}{{LN}K} \sum_{m=1}^{{LN}} \sum_{k=1}^{K} \beta_{m,k}
\end{equation}
This setting ensures that only channels with substantial gains are considered for each user, while those with lower gains are disregarded. Although it is dynamic based on the related AP and UE situations, this method does not account for instantaneous channel variations leading to sub-optimal spectral efficiency outcomes \cite{rscf}.
\subsection{BSR Based Clustering}

This method focuses on the sum-rate that can be achieved between an AP and UEs while ensuring reliable communication as a result of consideration of various factors such as the actual channel gain, the channel estimation error, and noise. Considering the downlink signal model from AP $m$ to UE $k$ as follows,
\begin{equation}
    y_{km}=\sqrt{\rho _{f}}\hat{{g}}_{km}^{T}\textbf{p}_{k}\mathbf{x}+\sqrt{\rho _{f}}\tilde{{g}}_{km}^{T}\textbf{p}_{k}\mathbf{x}+w_{k},
\end{equation}
where $\hat{{g}}_{km}$ represents the true channel gain between AP $m$ to UE $k$, $\tilde{{g}}_{km}$ is the estimation error of the channel gain, $\textbf{p}_{k}$ represents the precoder to UE $k$, and Gaussian signaling assumption, the rate from AP $m$ to UE $k$ is given by
\begin{equation}\label{eq:SRkm}
    {SR_{km} = \log_{2}\left (  1+\frac{\sqrt{\rho _{f}}\left | \hat{g}_{km} \right |^{2}\textbf{p}_{k}^{H}\textbf{p}_{k}}{\sqrt{\rho _{f}}\left | \tilde{g}_{km} \right |^{2}\textbf{p}_{k}^{H}\textbf{p}_{k}+\sigma _{w}^{2}}\right )}.
\end{equation}
To optimize AP selection, the approach considers the average rates between all UEs and APs. An average rate, $\alpha_{src}$, is defined as follows to ensure effective coverage and service quality:
\begin{equation} \label{alpha-asr}
    \alpha _{src}= \frac{1}{KM}\sum_{k=1}^{K}\sum_{m=1}^{M}SR_{km}
\end{equation}
This average rate sets the threshold for AP selection. APs that meet or exceed this rate form a cluster serving a specific UE. If no AP meets this threshold for a UE, the one providing the highest rate will serve, even if it falls below $\alpha_{src}$. The resulting AP cluster for UE $k$ is defined as:
\begin{equation} \label{ASR}
    U_{k,{bsr}}=\left\{ m \ : \ SR_{km}\ge \alpha _{src} \ \right\}\cup \left\{\underset{m}{\text{argmax}} \ SR_{km} \right\}.
\end{equation}
This strategy ensures that UEs are served by the APs capable of providing the highest information rates, prioritizing these rates over traditional metrics. Then a constraint is imposed to increase the minimum number of APs serving each UE. The detailed procedural steps of the proposed BSR technique are outlined in \cite{mashdour2024clustering}. 
 
Indeed, the BSR technique aims for the best possible performance based on the current conditions of the network and has the potential to offer better spectral efficiency.

As analyzed in \cite{mashdour2024clustering}, under conditions of high noise variance, relying solely on the LSF criterion for selecting APs may not lead to the best sum-rate performance. In contrast, the BSR method is more dynamic and tends to choose APs that provide superior links to UEs, thereby significantly enhancing overall system performance.


\section{Learning Based Clustering Schemes}

In this section, we consider the problem of AP selection by UEs as a multi-label classification problem. The multi-label classification problem is a variant of the classification problem where multiple nonexclusive labels may be assigned to each instance. Multi-label classification is a generalization of multi-class classification, which is the single-label problem of categorizing instances into precisely one of several (greater than or equal to two) classes. 
\subsection{Centeralized Approach}
We train a centralized fully connected feedforward deep neural network to do the AP selection. Therefore, this DNN uses network-wide information to do the AP selection in a user-centric manner and should be implemented at the CPU. 
Each training sample is designed in a user-centric manner, which means for a given UE $k$, the feature vector consists of the LSF coefficients between UE $k$ and all APs plus the UE exact location coordinates $\{x_{k}, y_{k}\colon \forall k\}$. Since for any specific UE the LSF coefficient for all $N$ antennas of AP $l$ is similar, we can denote the LSF coefficient between UEs and APs
by $\{\beta_{lk}\colon \forall k,l\}$ . Therefore the input matrix in LSF based case is designed as follows:

\begin{equation}
\mathbf{X}^{LSF} = \begin{bmatrix}
    \beta_{1,1} &  \cdots &  \beta_{L,1} &  x_1 &  y_1  \\
    \beta_{1,2} & \cdots & \beta_{L,2}  & x_2 & y_2 \\
    \vdots & \ddots & \vdots & \vdots & \vdots \\
    \beta_{1,K} & \cdots & \beta_{L,K} &  x_K & y_K
\end{bmatrix}    
\end{equation}
LSF coefficients can capture the main features of
propagation channels and interference and can
be easily measured in practice based on the received signal strength. The geographical location information of UEs is used to improve the performance of our model since they already capture the main feature of propagation
channels and interference in the network~\cite{sanguinetti2018deep}. In the case of BSR-based AP selection, the input matrix is designed as follows: 

\begin{equation}
\mathbf{X}^{BSR} = \begin{bmatrix}
    SR_{11} &  \cdots &  SR_{1M} &  x_1 &  y_1   \\
    SR_{21} & \cdots & SR_{2M}  & x_2 & y_2 \\
    \vdots & \ddots & \vdots & \vdots & \vdots \\
    SR_{K1} & \cdots & SR_{KM} &  x_K & y_K
\end{bmatrix}    
\end{equation}

The DNN then
attempts to learn the unknown mapping between the
input matrix and the final link status $\{a_{km}^*\colon \forall k, m\}$. For training, two separate training sets containing a large number of training pairs $\{\beta_{lk}, a_{km}^{*}: \forall l,m, k\}$ and $\{SR_{km}, a_{ km}^{*}: \forall m, k\}$, are generated using the LSF and BSR clustering approaches. We define the ground truth output as follows:

\begin{equation}
\mathbf{A}^{*} = \begin{bmatrix}
     a_{11}^{*} & \cdots & a_{1M}^{*}\\
    a_{21}^{*} & \cdots & a_{2M}^{*} \\
    \vdots & \ddots & \vdots  \\
    a_{K1}^{*} & \cdots & a_{KM}^{*}
    \end{bmatrix} 
\end{equation}
The DNN is trained to minimize the following multi-class cross-entropy loss:

\begin{equation}
    CE= - \mathbf{A}^{*} \log(\mathbf{A}^{DNN})
    \label{CE}
\end{equation}

The loss function described in~\eqref{CE} is averaged
over the training samples, and the DNNs aim to minimize this loss function. The training process contains the updating of weights
between neurons as well as the bias in each layer. The Adam
optimizer is employed and the learning rate is set to 0.001. The batch size and number of epochs are chosen using a trial-and-error method.

\subsection{Distributed Approach}

\begin{figure}
    \centering
    \includegraphics[width=.8\linewidth]{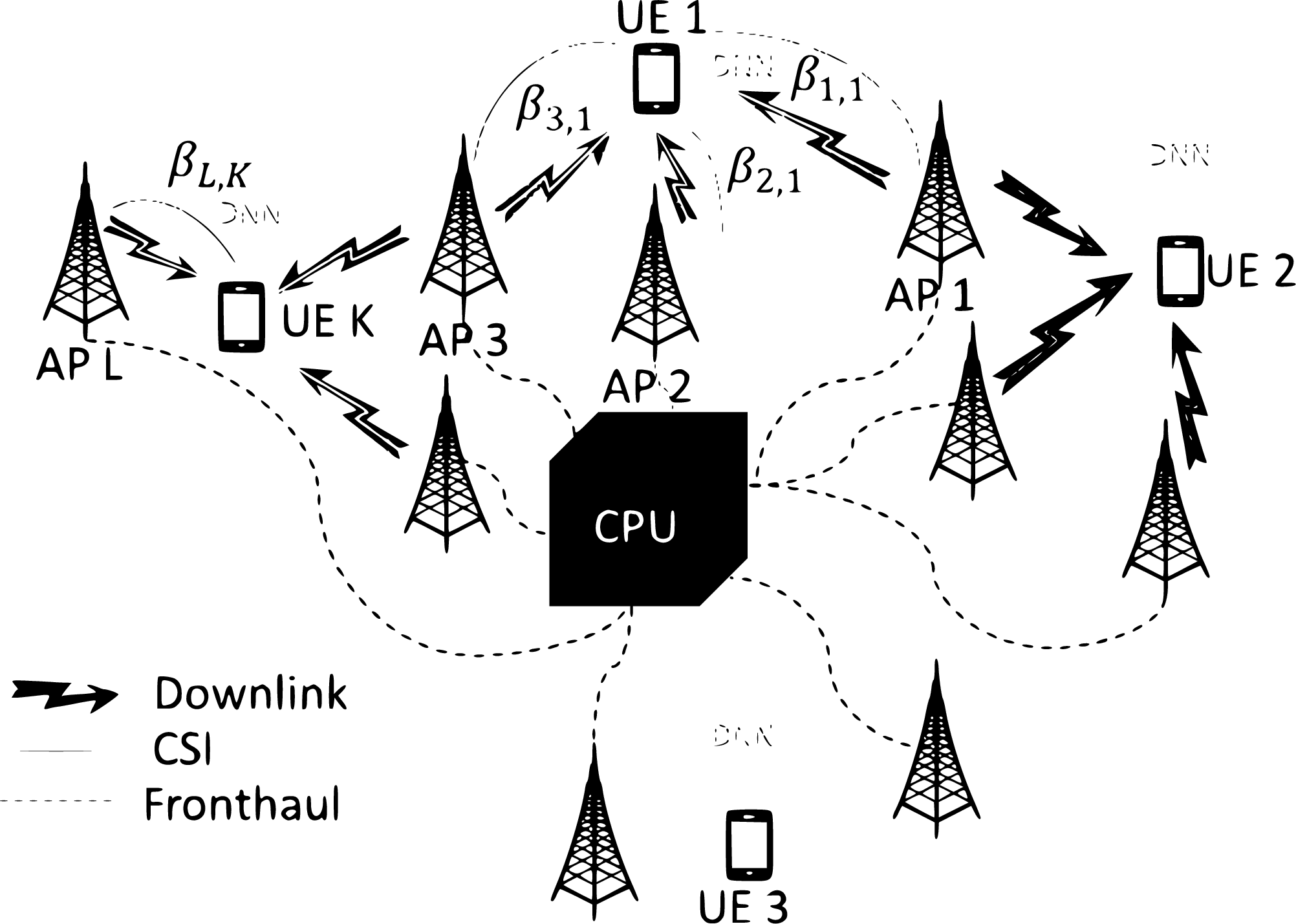}
    \vspace{-1em}
    \caption{UCCF clustered network with distributed DNNs to manage AP selection.}
    \vspace{-1em}
    \label{distributed}
\end{figure}

Targeting a solution that is both less computationally intensive and distributed for scalability, we train a DNN model for each UE, as illustrated in Fig.~\ref{distributed}. In this approach, the training of each DNN is limited to the local information of the corresponding UE and not the whole network-wise information of the previous approach. For the labeled outputs, however, we utilize the same network-wide solution given by LSF or BSR algorithms. Hence, for each UE $k$, the DNN
attempts to learn the unknown mapping between the locally
available coefficients $\{\beta_{lk}\colon \forall l\}$ plus the UEs exact location coordinates $\{x_{k}, y_{k}\}$ and the optimal active links corresponding to that UE $\{a_{km}^*\colon \forall m\}$. The input and output vectors for each UE $k$ are defined as follows:

\begin{equation}
\mathbf{x}^{LSF}_k = \begin{bmatrix}
    \beta_{1,k},  \cdots,  \beta_{L,k},  x_k,  y_k  \\
\end{bmatrix}    
\end{equation}

\begin{equation}
\mathbf{x}^{BSR}_k = \begin{bmatrix}
    SR_{k1},  \cdots,  SR_{kM},  x_k,  y_k  \\
\end{bmatrix}    
\end{equation}

\begin{equation}
\mathbf{a}^{*}_k = \begin{bmatrix} a_{k1}^{*}, \cdots, a_{kM}^{*}  
\end{bmatrix}
\end{equation}

This eliminates the need for extensive transmission of LSF coefficients to the CPU, unlike in a centralized setup, enabling a scalable network operation. Additionally, there is a notable reduction in the number of parameters requiring training for each DNN.

\subsection{Computational Complexity Analysis}

The computational complexity of this approach is mostly
in the data generation phase for training the DNN, where the
actual iterative AP selection algorithms have to be implemented.
On the other hand, for a DNN with $T$ layers such that layer $i$
has $T_i$ neurons, the required number of real multiplications and additions is each $T_iT_{i-1}$, at layer $i$, $i = 1, \cdots, T$. Moreover, $\Sigma_{i=1}^{T} T_i
$ activation functions need to be evaluated in total.

\section{Numerical Results and Analysis}
In this section, we evaluate the performance of our centralized and distributed DNN-based AP selection solutions compared to the LSF and BSR algorithms. We consider a cell-free network consisting of $L=16$ APs each equipped with $N=4$ antennas in an area of $400\mathrm{m} \times 400\mathrm{m}$. In each simulation instance, $K=32$ UEs are distributed randomly and uniformly within the area of interest. We generate 6000 simulation instances including six different SNR levels. 
The structure of the DNN  models for centralized and distributed approaches are shown in Tables~\ref{layout1} and~\ref{layout2}, respectively. We use the same architecture for both centralized and distributed models. However, as we see the size of the input layer (LSF/BSR) is different in centralized and distributed approaches.
\setlength{\tabcolsep}{1.5pt}
\begin{table}
    \centering
    \caption{Layout of the centralized DNN for AP selection (LSF/BSR).
Parameters to be trained: 293088/342240}
    \begin{tabular}{cccc}
    & Size  & Parameters & Activation Functions\\
    \hline
       Input  & $K(L+2)/K(M+2)$ & - & - \\
       Layer 1 (Dense)  & 32 & 18464/67616 & linear\\
        Layer 2 (Dense) & 64 & 2112 & relu\\
       Layer 3 (Dense)  & 128 & 8320 & relu\\
       Layer 4 (Dense)  & $KM$ & 264192 & sigmoid\\
       \hline
    \end{tabular}

    \label{layout1}
\end{table}

\begin{table}
    \centering
    \caption{Layout of the distributed DNN for AP selection (LSF/BSR).
Parameters to be trained: 19296/20832}
    \begin{tabular}{cccc}
    & Size  & Parameters & Activation Functions\\
    \hline
       Input  & $L+2/M+2$ & - & - \\
       Layer 1 (Dense)  & 32 & 608/2144 & linear\\
        Layer 2 (Dense) & 64 & 2112 & relu\\
       Layer 3 (Dense)  & 128 & 8320 & relu\\
       Layer 4 (Dense)  & $M$ & 8256 & sigmoid\\
       \hline
    \end{tabular}
    \label{layout2}
\end{table}

In order to evaluate our models, we use 5-fold cross-validation. This means that our data is randomly divided into 5 groups of equal size. The first fold is treated as a validation set, and the model is fit on the remaining 4 folds. Therefore, the size of the test set is 1200. However, the results are averaged over 6000 UE distributions. Each value of the output vector is considered 1 if the predicted value is more than 0.5 and 0 otherwise. The performance of the centralized and distributed learning model in terms of true positive rate (TPR) and true negative rate (TNR) are illustrated in Table~\ref{confusion} and Table~\ref{confusion 2}, respectively.

\begin{table}
    \centering
        \caption{Average TPR and TNR for centralized AP selection approaches}
    \begin{tabular}{ccc}
       Algorithm  & TPR(\%) & TNR(\%)\\
       \hline
       DNN-LSF  & 88.6 & 98.9 \\
       DNN-BSR  & 84.9 & 97.6 \\
       \hline
    \end{tabular}
    \label{confusion}
\end{table}

\begin{table}
    \centering
        \caption{Average TPR and TNR for distributed AP selection approaches}
    \begin{tabular}{ccc}
       Algorithm  & TPR(\%) & TNR(\%)\\
       \hline
       DNN-LSF  & 80.2 & 99.3 \\
       DNN-BSR  & 75.7 & 97.9 \\
       \hline
    \end{tabular}
    \label{confusion 2}
\end{table}

\begin{figure}
    \centering
    \includegraphics[width=0.7\linewidth]{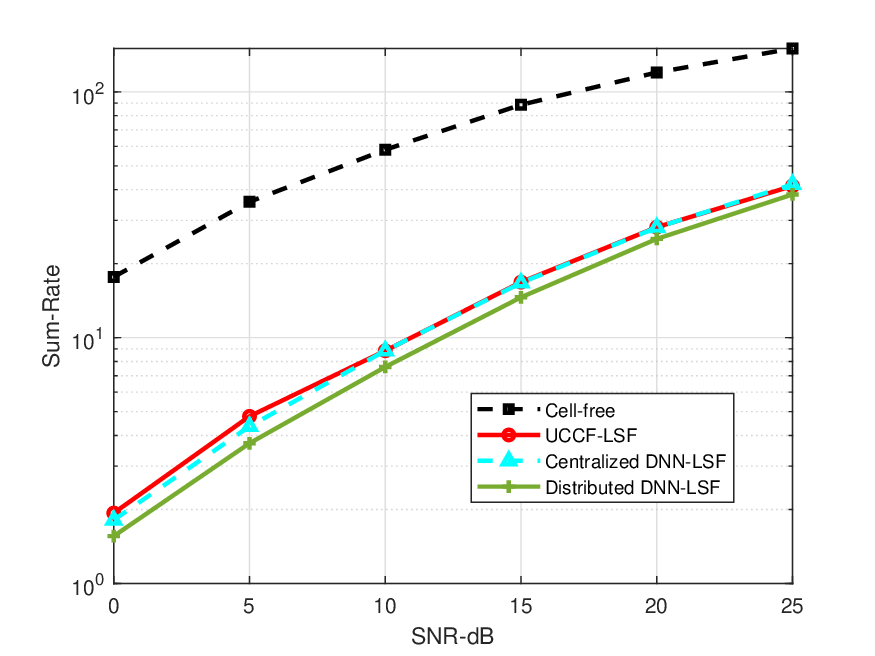}
    \caption{
    Sum-rate comparison of learning-based AP selection methods trained by LSF coefficients for $K=32$ and $M=64$.}
    \label{sumrate comparison lsf}
\end{figure}

\begin{figure}
    \centering
    \includegraphics[width=0.7\linewidth]{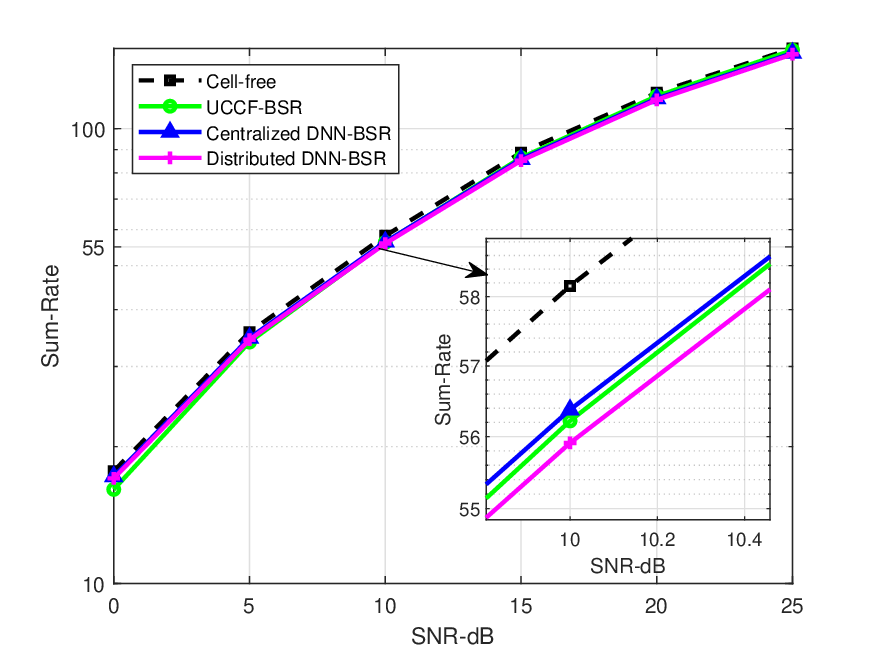}
    \caption{
    Sum-rate comparison of learning-based AP selection methods trained by BSR coefficients for $K=32$ and $M=64$.}
    \label{sumrate comparison bsr}
\end{figure}

The performance of DNN-based AP selection schemes is compared to LSF or BSR algorithms in terms of sum-rate and the results are shown in Figs.~\ref{sumrate comparison lsf} and~\ref{sumrate comparison bsr} for different levels of SNR. It can be seen that the learning-based solutions can follow the UCCF-LSF and UCCF-BSR approaches almost perfectly. The distributed approach performs slightly poorer than the centralized approach. This is expected as the distributed DNNs are utilizing only the local information. 

In Table~\ref{runtime}, we recorded the
average per UE run-time for AP selection using LSF and BSR-based clustering algorithms (UCCF-LSF and UCCF-BSR) as well as the DNN models for both approaches (DNN-LSF and DNN-BSR). As we can see, the required time for a UE to perform AP selection is 18 to 26 times faster in the case of using DNNs in the inference phase. Please also note that a learning-based solution has also the benefit of resilience to the UE mobility compared to the non-learning approaches that need to do all the computations from scratch as soon as the UE locations change. In such cases, the learning-based approach just needs to repeat the inference phase without the need for retraining.
\begin{table}
    \centering
        \caption{Comparison of per UE AP selection run time in milliseconds}
    \begin{tabular}{cc}
       Algorithm  & Time (ms)\\
       \hline
       UCCF-LSF  & 14.6\\
       UCCF-BSR & 21.2\\
       Centralized DNN-LSF  & 0.8\\
       Centralized DNN-BSR  & 0.8 \\
       Distributed DNN-LSF  & $608\times10^{-3}$\\
       Distributed DNN-BSR  & $608\times10^{-3}$\\
       \hline
    \end{tabular}
    \label{runtime}
\end{table}
\vspace{-5mm}
\section{Conclusion and Future Works}
In this paper, centralized and distributed learning-based solutions are developed for AP selection in UCCF massive MIMO systems. In the centralized approach, the problem is formulated in a user-centric manner with network-wide information. Two training sets were generated from two heuristic AP selection algorithms and the DNN is trained on these datasets to learn the unknown mapping between the LSF/BSR coefficients and the final link status between all UEs and APs in a multi-label classification task. In the distributed version, 32 DNNs are trained with the local information of each UE. The results demonstrate that the learning-based solutions closely follow the sum-rate performance curves of the heuristic algorithms while significantly reducing computation time. This efficiency makes the learning-based approach particularly suitable for real-time scenarios, where the dynamic nature of user movements can be effectively managed using the DNN model during the inference phase. It is worth noting that the performance of the UCCF clustered network based on DNN-BSR closely approximates the ideal performance of a CF network. Our findings highlight the potential of learning-based models to enhance the scalability, adaptability, and performance of UCCF massive MIMO systems, making them a promising option for future wireless networks.
In the future, we aim to extend our learning-based methods to perform a comprehensive dynamic resource allocation framework for UCCF networks. This framework usually includes precoder design, multiuser scheduling, and power allocation in addition to AP selection. We also aim to apply our learning-based approaches to a robust resource allocation framework~\cite{mashdour2024robust,rrscf} that mitigates the effects of imperfect CSI on network performance, ensuring that the system is resilient to errors in channel estimation. Our approach may utilize supervised learning, leveraging
previously developed analytical and heuristic solutions, or employ reinforcement learning to achieve enhanced network performance. 
\section*{Acknowledgement}
The authors acknowledge the financial support by the Federal Ministry of Education and Research of Germany in the project “Open6GHub” (grant number: 16KISK012).
\vspace{-3mm}
\bibliographystyle{IEEEbib}
\bibliography{Refs}

\end{document}